\documentclass[aps,prb,preprint]{revtex4}

\usepackage{graphicx}
\usepackage[utf8]{inputenc}

\begin{document}

\title{Spin pumping and interlayer exchange coupling through palladium}

\author{D. L. R. Santos}
\affiliation{Instituto de F\'isica, Universidade Federal Fluminense, 24210-346 Niter\'oi, RJ, Brazil.}
\author{P. Venezuela}
\affiliation{Instituto de F\'isica, Universidade Federal Fluminense, 24210-346 Niter\'oi, RJ, Brazil.}
\author{R. B. Muniz}
\affiliation{Instituto de F\'isica, Universidade Federal Fluminense, 24210-346 Niter\'oi, RJ, Brazil.}
\author{A. T. Costa}
\email{antc@if.uff.br}
\affiliation{Instituto de F\'isica, Universidade Federal Fluminense, 24210-346 Niter\'oi, RJ, Brazil.}

\begin{abstract}
The magnetic behaviour of ultrathin ferromagnetic films deposited on substrates is strongly
affected by the properties of the substrate. We investigate the spin pumping rate, 
interlayer exchange coupling and dynamic exchange coupling between ultrathin ferromagnetic 
films through palladium, a non-magnetic substrate that displays strong Stoner enhancement. 
We find that the interlayer exchange coupling, both in the static and dynamic versions,
is qualitatively affected by the substrate's Stoner enhancement. For instance, the oscillatory
behavior that is a hallmark property of the RKKY exchange coupling is strongly suppressed by Stoner 
enhancement. Although the spin pumping rate of ferromagnetic films atop palladium is only
mildly changed by Stoner enhancement the change is large enough to be detected experimentally.
The qualitative aspects of our results for palladium are expected to remain valid for any non-magnetic 
substrate where Coulomb repulsion is large. 
\end{abstract}

\maketitle

\section{Introduction}

Palladium is a fascinating material. In bulk form it is non-magnetic, but it is on the brink of becoming ferromagnetic: 
the product of its paramagnetic density of states at the Fermi level $\rho(E_F)$ by the intra-atomic effective Coulomb interaction $U$ is very close to the critical value of 1. The bulk conductivity as a function of temperature displays a $T^2$ contribution due to scattering by spin fluctuations that dominates the contribution from scattering by phonons at low $T < 10$ K. Adding Ni impurities to bulk Pd in concentrations smaller than 2\% increase dramatically the effect of spin fluctuations on electronic transport: the coefficient of the $T^2$ term in the conductivity is amplified by a factor of ten.~\cite{LedererMills_NiPd}

These remarkable properties indicate that Pd is a non-magnetic material where the effects of electron-electron interactions are strong enough to be observed experimentally. In fact, direct observation of paramagnons in bulk Pd have been reported very recently.\cite{paramagnon_observation}

Palladium is still a challenge to our most successful theory for the electronic structure of materials. Some implementations of DFT predict bulk Pd to be magnetic and it is usually necessary to tune the lattice constant used in the calculations to obtain non-magnetic bulk Pd.\cite{PhysRevLett.96.079701}

The importance of non-magnetic substrates in determining the spin dynamics of 
adsorbed ultrathin films and adatoms has been demonstrated unequivocally in recent years. 
Excitation energies and lifetimes may be strongly affected by the hybridisation between 
magnetic entities' and substrates' electronic states. Nice examples are provided by ultrathin Fe layers
on W(110), where the strong spin-orbit coupling in W induces a large anisotropy and a strong Dzyaloshinskii-Moryia
coupling between Fe magnetic moments,\cite{FeW110_SOC_Theory,FeW110_SOC_Experiment_1} and by Fe adatoms on Cu(111)~\cite{FeCu111_STS} and Ag(111),\cite{FeAg111_STS}
where the hybridisation with the substrate strongly
dampens the spin excitations and produces large shifts in their energies compared to isolated atoms.

The effects of Pd's large Stoner enhancement have been thoroughly studied in the past. However, there seems to be
relatively few studies exploring enhancement effects on interlayer exchange coupling or spin pumping. 
One of these studies~\cite{heinrich_Pd} have shown
that Pd may serve as a good sink for spin currents emanated from ferromagnetic materials. There it is suggested that
the absorption of spin current is facilitated by the enhanced spin fluctuations in Pd. 

Interlayer exchange coupling (IEC) has received enormous attention during the 1990's, specially due to
its relation to the discovery of giant magnetoresistance and its subsequent application to read heads 
in information storage devices. Besides its huge technological importance, IEC is also a very intriguing phenomenon. 
Its basic physics can be easily understood in terms of very simple models for the electronic structures of the
materials involved, but there are many subtle behaviours that can only be fully accounted for if
one employs realistic band structure calculations. Most realistic calculations of IEC in layered
systems employ a non-interacting description of the spacer material. While this is certainly appropriate
for free-electron-like metals as Cu, Ag and Au, it is less so for transition metals. Extreme
examples would be Pd and Pt which, although non-magnetic in bulk form, exhibit very large Stoner 
enhancement. To the best of our knowledge, the influence of Coulomb interaction within the spacer
layer on the properties of IEC remains unexplored, except for the work reported in 
Ref.~\onlinecite{takahashi_FePdFe}. Takahashi studies the IEC between Fe layers separated by
Pd by mapping the interaction onto a classical Heisenberg hamiltonian and uses the static
RPA susceptibility to estimate the effect of Pd's Stoner enhancement.

Another form of interlayer coupling has gained notoriety in recent years due to seminal works by
Slonczewsky~\cite{Slonczewski1999L261} and Berger.\cite{berger_PhysRevB.54.9353} The idea is
that a ferromagnet in contact with a non-magnetic metallic medium will, when perturbed from its equilibrium 
configuration, inject a spin
current into the medium, that can be absorbed by a second ferromagnet in contact with the same 
medium.\cite{Tserkovnyak_RMP2005}
This is very attractive from the technological point of view, since signal transmission by pure
spin currents would be much less affected by the dissipation mechanisms that strongly affect
charge currents. That said, there are other mechanisms that can influence pure spin currents and
one of them is the Coulomb repulsion between electrons within the metallic medium.

In this paper we will present the results of our investigations concerning the effects of strong
Stoner enhancement on the properties of the ``traditional'', RKKY-like interlayer exchange coupling
and on the dynamical coupling promoted by the emission and absorption of pure spin currents by 
ferromagnetic films. We employed realistic electronic structure calculations to describe both the ground 
state and the spin dynamics of layered systems composed of several atomic layers of Co or Ni 
deposited on Pd substrates. The formalism we used has been described in great detail in previous
publications.\cite{SW_1band,SW_Fe110_old1,SW_Fe110_old0,SpinPumpingTheory,DynCouplGeneral,DynCouplCN} 
Thus, in section~\ref{theorysection}  we remind the reader of the main features of our theoretical
approach and establish the notation that will be used throughout the paper. In section~\ref{resultsection}
we present and discuss the results of our numerical calculations and in section~\ref{concludingsection}
we offer our final remarks.

\section{Theoretical Framework}
\label{theorysection}

\subsection{Electronic structure and self-consistent ground state}
\label{GSSection}

We describe the electronic structure of the system using a multi-orbital extension of the Hubbard model,
\begin{equation}
H = \sum_{ll'}\sum_{\mu\nu}\sum_\sigma t^{\mu\nu}_{ll'}a^\dagger_{l\mu\sigma}a_{l'\nu\sigma} +
\sum_l\sum_{\sigma\sigma'}\sum_{\mu\nu\mu'\nu'}U^{\mu\nu\mu'\nu'}_l 
a^\dagger_{l\mu\sigma}a^\dagger_{l\nu\sigma'}a_{l\nu'\sigma'}a_{l\mu'\sigma}+
\frac{g\mu_B B_z}{2}\sum_{l,\mu}(a^\dagger_{l\mu\uparrow}a_{l\mu\uparrow}-
a^\dagger_{l\mu\downarrow}a_{l\mu\downarrow}) .
\label{multiorbhubbard}
\end{equation} 
where $a^\dagger_{l\mu\sigma}$ creates an electronic state in atomic orbital $\mu$ at lattice site 
$l$ with spin $\sigma$. $t^{\mu\nu}_{ll'}$ are hopping matrix elements extracted from DFT-based
calculations and $U^{\mu\nu\mu'\nu'}_l$ are the matrix elements of the effective on-site Coulomb
interaction. This model provides magnetizations, local densities of states and local 
occupancies in excellent agreement with DFT-based calculations. The last term in Eq.~\ref{multiorbhubbard}
is the Zeeman energy corresponding to a uniform magnetic field of intensity $B_z$ applied to the sample
along the $z$ axis. It is
essential to the description of ferromagnetic resonance experiments. The constants $g$ and $\mu_B$
are the electron's gyromagnetic factor and the Bohr magneton, respectively.

Here we address the self-consistent ground state (SCGS) of ferromagnetic films of Ni or Co adsorbed to Pd substrates.
We employed bulk tight-binding parameters to describe the electronic structures of both magnetic film and 
Pd substrates. This is not ideal but we believe it is not critical either, since the effects we want to
explore are not related to details of the materials' band structures but to a very robust feature of bulk Pd, 
namely its high Stoner enhancement. In order to determine the SCGS a mean-field approximation is adopted in which
the interaction term in Eq.~\ref{multiorbhubbard} is replaced by 
\begin{equation}
\sum_{l}\sum_{\mu\in d}\sum_{\sigma}V^0_la^\dagger_{l\mu\sigma}a_{l\mu\sigma}-
\sum_{l}\sum_{\mu\in d}\frac{U_lm}{2}(a^\dagger_{l\mu\uparrow}a_{l\mu\uparrow} - a^\dagger_{l\mu\downarrow}a_{l\mu\downarrow})
+
\frac{g\mu_B B_z}{2}\sum_{l,\mu}(a^\dagger_{l\mu\uparrow}a_{l\mu\uparrow}-
a^\dagger_{l\mu\downarrow}a_{l\mu\downarrow}) .
\label{meanfield}
\end{equation}
The on-site potential $V^0_l$ at each magnetic site is adjusted in order to guarantee local charge neutrality.
The fact that the self-consistent mean field depends only on the magnetization is a consequence of the
simple form we adopt for the matrix elements of the Coulomb interaction,
\begin{equation}
U_l^{\mu\nu\mu'\nu'} = U_l\delta_{\mu\nu'}\delta_{\nu\mu'}
\label{LW}
\end{equation}
A more complex parametrization depending on three parameters has been tested in several cases and shown to yield only very small quantitative differences when compared to the much simpler approximation
described by Eq.~\ref{LW}.\cite{SW_Fe110_old0} 

\begin{figure}
\includegraphics[width=0.8\textwidth]{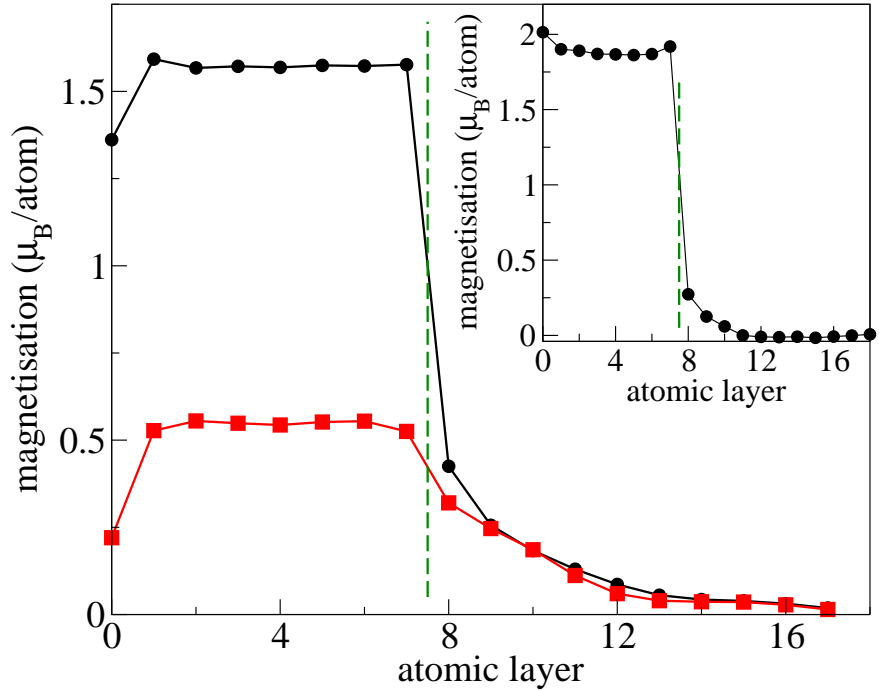}
\caption{Magnetic moment per atom at each atomic layer from the surface of the film (layer 0) to
the 10th layer of Pd (layer 17). The green dashed line marks the interface between the
ferromagnetic film (Co, black solid circles, and Ni, red solid squares) and the Pd substrate.
The inset shows the results of a DFT-based calculation for 8Co/Pd(001).}
\label{magneticmomentsCoPd}
\end{figure}

The famous Stoner criterion for the appearance of local magnetic moments in metals establishes 
that the ground state of the metal will be spin polarized provided that the product
of the effective Coulomb repulsion intensity $U$ by the paramagnetic density of states per spin 
at the Fermi level $\rho^0(E_F)$ exceeds unity. In the 3d transition metals Fe, Co and Ni
the criterion in easily satisfied. In noble metals such as Cu, Au and Ag, $U\rho^0(E_F)$
is well below 1. In Pd, however, specific heat and uniform spin susceptibility measurements indicate that
$U\rho^0(E_F)\sim 0.9$.~\cite{schrieffer1964,DoniachEngelsberg1966}
This means that, while the ground
state of bulk Pd has negligible spin polarization, correlation effects should not be
neglected. There are many strong evidences of the importance of correlation effects in Pd,
some of which have been mentioned in the introduction. One strong evidence that palladium
is on the verge of becoming ferromagnetic is the giant polarisation cloud produced by Fe 
impurities added in very small concentrations to bulk Pd.\cite{GiantMM_Pd}

\begin{figure}
\includegraphics[width=0.8\textwidth]{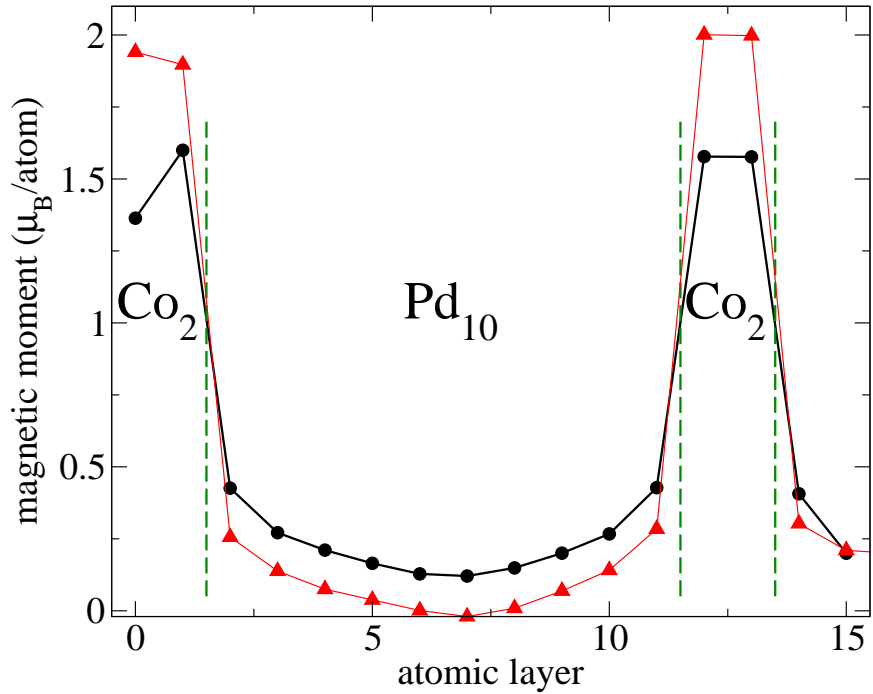}
\caption{Magnetic moment per atom at each atomic layer of Co$_2$/Pd$_{10}$/Co$_2$/Pd(001) (solid circles).  
The vertical dashed lines mark the interfaces between Co and Pd. Only the two layers of the substrate closest to the second Co
film are shown. The triangles represent the magnetizations obtained from a DFT calculation on the 
same system (Co$_2$/Pd$_{10}$/Co$_2$/Pd(001)).}
\label{m_trilayer}
\end{figure}

We first consider ferromagnetic films of Co or Ni deposited on semi-infinite Pd substrates.
As shown in Fig.~\ref{magneticmomentsCoPd}, the magnetic moments in the substrate are
far from negligible, specially within the atomic layers close to the interface. For the Ni/Pd(001)
system the magnetisation of the Pd layer closest to the interface is comparable to the average
magnetisation of the Ni layers. In Fig.~\ref{m_trilayer} we show the calculated magnetic moments
of a Co$_2$/Pd$_{10}$/Co$_2$ trilayer deposited on the surface of semi-infinite Pd(001). These
results are similar to the ones depicted in Fig.~\ref{magneticmomentsCoPd}.

In order to verify if the ground state magnetizations obtained within our tight-binding model
are compatible to the predictions of density functional theory we performed an ab-initio calculation
using the VASP package.\cite{vasp1,vasp2} We employed a functional based on the local density approximation 
(LDA) for exchange-correlation effects.\cite{perdewzunger} LDA was used because it leads
to a better description of the magnetic properties of Pd, compared to generalized gradient approximation calculations.\cite{PhysRevLett.96.079701} The interaction between the valence electrons and the ionic cores was taken in account 
by means of the Projector Augmented-Wave Method (PAW).\cite{paw_method} The single particle wavefunctions were 
expanded in a plane wave basis up to an energy cutoff of 350 eV. To sample de Brillouin Zone the Monkhorst Pack 
method was used with 12x12x1 points grids.\cite{MonkhorstPack} Large unit cells with vaccum regions 4$a_0$ tick were employed in order to simulate the surfaces, where $a_0$ = 3.89~\AA\  is the experimental lattice parameter of Pd.

\subsection{Spin Dynamics}

The spin dynamics is investigated, within the linear response regime, through the transverse spin
susceptibility matrix
\begin{equation}
\chi_{\mu\nu}(\vec{R}_l,\vec{R}_{l'};t) = -i\theta(t)\left\langle\left[ S^+_\mu(\vec{R}_l;t),S^-_\nu(\vec{R}_{l'};0)
\right]\right\rangle .
\label{eqchipm}
\end{equation}
$\mu,\nu$ are orbital indices, $\vec{R}_l$ is the position of lattice site $l$, $S^+_\mu(\vec{R}_l)=a^\dagger_{l\mu\uparrow}a_{l\mu\downarrow}$ and $S^-_\mu(\vec{R}_l)=[S^+_\mu(\vec{R}_l)]^\dagger$. From the (time) Fourier transform 
of $\chi_{\mu\nu}(\vec{R}_l,\vec{R}_{l'};t)$ it is possible to extract the response to transverse magnetic fields and
the magnon density of states. For the film geometry it is convenient to work in a mixed basis of Bloch states
within each atomic layer and localised states in the direction perpendicular to the layers,
\begin{equation}
|\vec{k}_\parallel;l;\mu\rangle = \frac{1}{N_\parallel}\sum_{\vec{R}_\parallel}e^{i\vec{k}_\parallel\cdot\vec{R}_\parallel}|\vec{R}_\parallel;l;\mu\rangle .
\end{equation}
$|\vec{R}_\parallel;l;\mu\rangle$ is the state vector associated with an atomic orbital $\mu$ at site 
$\vec{R}_\parallel$ within atomic layer $l$. For systems with translation symmetry within the atomic layers,
the susceptibility defined by Eq.~\ref{eqchipm} is a function of the relative position in the
direction parallel to the atomic layers,
\begin{equation}
\chi_{\mu\nu}(\vec{R}_l,\vec{R}_{l'};t) = \chi_{\mu\nu}(\vec{R}_\parallel-\vec{R}^{'}_\parallel;l,l';t)  .
\end{equation}
It is most useful to calculate the Fourier transform of Eq.~\ref{eqchipm} within the atomic layers, together with the Fourier transform from time to frequency (or energy)
domain. The object we are interested in is, thus,
\begin{equation}
\chi_{\mu\nu}(\vec{Q}_\parallel;l,l';\Omega) = 
\frac{1}{N_\parallel}\sum_{\vec{R}_\parallel}e^{i\vec{Q}_\parallel\cdot\vec{R}_\parallel}\int d\omega e^{i\Omega t}
\chi_{\mu\nu}(\vec{R}_\parallel;l,l';t) .
\label{qparchipm}
\end{equation}
It can be easily shown that this is precisely the response function of the system to a 
magnetic field circularly polarized in the direction perpendicular to the equilibrium magnetization,
whose space-time dependence in the direction parallel to the layers is a plane wave with
wave vector $\vec{Q}_\parallel$ and frequency $\Omega/\hbar$. The spectral density associated with this
transverse susceptibility may also be interpreted as the partial density of states of magnons at layer $l$
with wave vector $\vec{Q}_\parallel$,
\begin{equation}
A_l(\vec{Q}_\parallel;\Omega) = -\frac{1}{\pi}\mathrm{Im}\sum_{\mu\nu}\chi_{\mu\nu}(\vec{Q}_\parallel;l,l;\Omega) .
\label{eqspectraldensity}
\end{equation}
From the peaks of $A_l(\vec{Q}_\parallel;\Omega)$ as a function of $\Omega$ for fixed values of $\vec{Q}_\parallel$
it is possible to extract the spin wave spectrum of the system. The particular case $\vec{Q}_\parallel=0$ is
specially interesting: it corresponds to the uniform excitation field applied to the sample in 
ferromagnetic resonance (FMR) experiments. Thus, we can calculate the spin wave spectra and the
FMR spectra of ultrathin magnetic films within the same formalism. Moreover, we have information
about the spectral intensities on a layer-by-layer basis, which allows us to study non-local
spin responses with relative ease.~\cite{SpinPumpingTheory}

\section{Results}
\label{resultsection}
\subsection{Spin Pumping}

We start by analysing the FMR spectra of several Co and Ni ultrathin films on Pd(001). The FMR
linewidth is directly related to the spin current pumped by the ferromagnetic film into the substrate. 
It has been argued, based on geometric reasoning, that the FMR linewidth should decrease as the 
inverse of the ferromagnetic film thickness $N$, due to the fact that spin pumping is an interface phenomenon. 
This argument, however, ignores the effects of quantum interferences within the ferromagnetic film and the 
fact that the electronic structure of the film depends on its thickness. This is specially
true for the ultrathin films (1 to $\sim 10$ atomic layers thick) we consider here. 

When the FMR precession is described phenomenologically through the Landau-Lifshitz-Gilbert
equation, as it very frequently is, the effective damping term usually takes the form 
\begin{equation}
\alpha\hat{m}\times\frac{d\hat{m}}{dt}
\end{equation}
where $\hat{m}$ is a unit vector parallel to the sample's magnetisation and $\alpha$ is the Gilbert 
constant, which embodies the damping due to all possible mechanisms. If one assumes spin pumping is the only 
active damping mechanism, $\alpha$  should be proportional to the ratio between the FMR linewidth $\Delta\Omega$ 
and the resonance frequency $\Omega_0$. We have chosen this ratio as the measure of 
spin pumping rate in our calculations. 

One virtue of our model calculations is that we can test several contributions to the damping rate separately. 
Pd has a large Stoner enhancement and, of course, its d bands are crossed by the Fermi energy. Thus we first
turn off Stoner enhancement by making the strength of the effective interaction in Pd equal to zero.
This amounts to making $U^{\mu\nu\mu'\nu'}_l=0$ in Eq.~\ref{multiorbhubbard} whenever the value of $l$ 
corresponds to a Pd site. 

\begin{figure}
\includegraphics[width=0.8\textwidth]{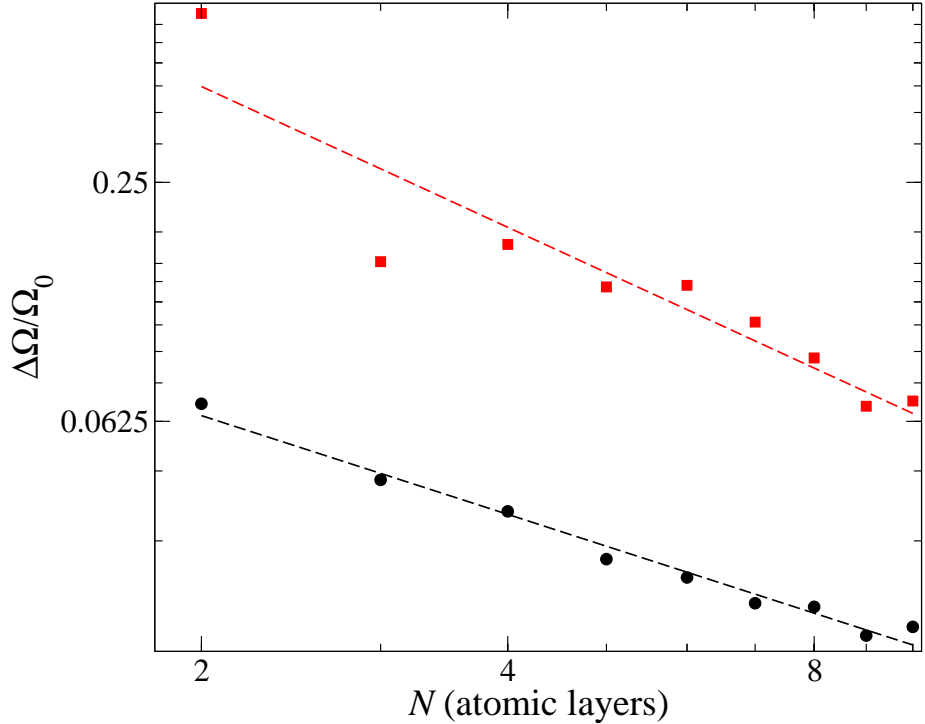}
\caption{Linewidths of the FMR spectra for $N$Co/Pd(001) (solid black circles) and $N$Ni/Pd(001) 
(solid red squares). Stoner enhancement in Pd has been turned off. The dashed lines are fittings to 
power laws $N^{-\alpha}$ with exponents $\alpha_\mathrm{Pd}\sim 0.83$ and $\alpha_\mathrm{Ni}\sim 1.2$.}
\label{FMR_UPd_0}
\end{figure}

The linewidths extracted from FMR calculations using such approach exhibit a clear power-law decrease  
as a function of $N$, as shown in Fig.~\ref{FMR_UPd_0}. The power law exponent is $\sim 0.83$ for Co and $\sim 0.74$ for Ni. 
Previous calculations on ultrathin Fe films on W,\cite{SpinPumpingTheory} which has negligible Stoner 
enhancement but whose d bands are crossed by the Fermi energy, also show a power law exponent close to 0.8.
When W was replaced by free-electron-like substrates such as Au and Ag, the exponent came out very close to one.
It is, however, difficult to attribute such deviations to general electronic structure features of the substrate.
In fact, as we have just seen, two different magnetic materials on the same substrate can lead to two very
different power-law behaviors.

\begin{figure}
\includegraphics[width=0.8\textwidth]{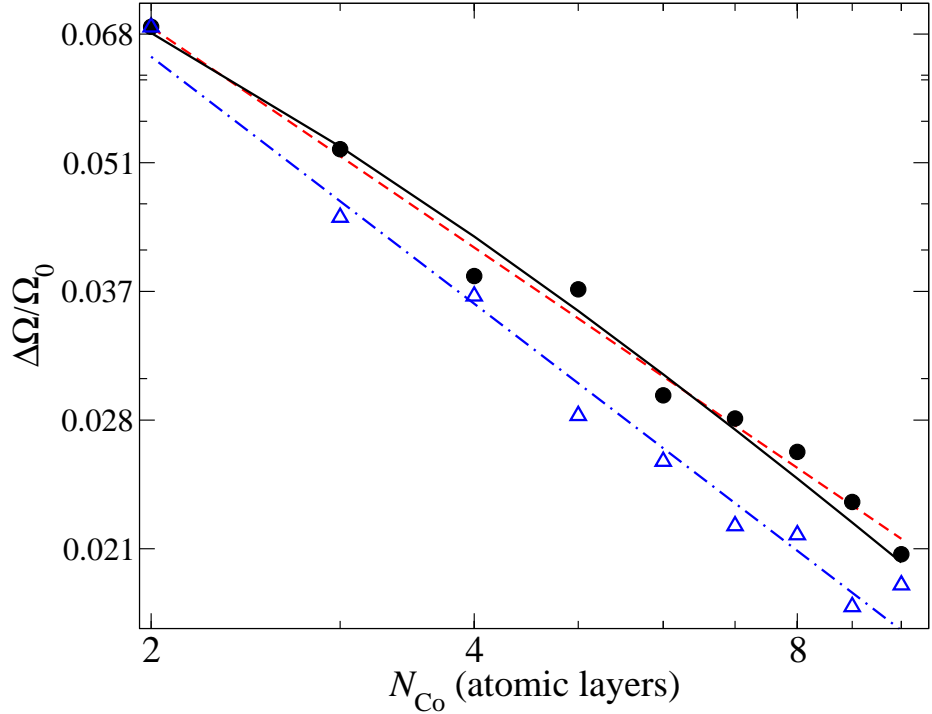}
\caption{Damping rate as a function of the Co film thickness for Stoner-enhanced Pd (solid circles) 
and non-enhanced Pd (open triangles) substrates. The solid line is a $1/(N+N_0)$
fitting of the results for the enhanced substrate (see text). The (red) dashed line 
is a $1/N^\alpha$ fitting of the results for the enhanced substrate, with $\alpha\approx 0.74$.
The dot-dashed line
is a $1/N^\alpha$ fitting of the results for the non-enhanced substrate with $\alpha\approx 0.83$ (also shown
in Fig.~\ref{FMR_UPd_0}).}
\label{FMR_CoPd}
\end{figure}

\begin{figure}
\includegraphics[width=0.8\textwidth]{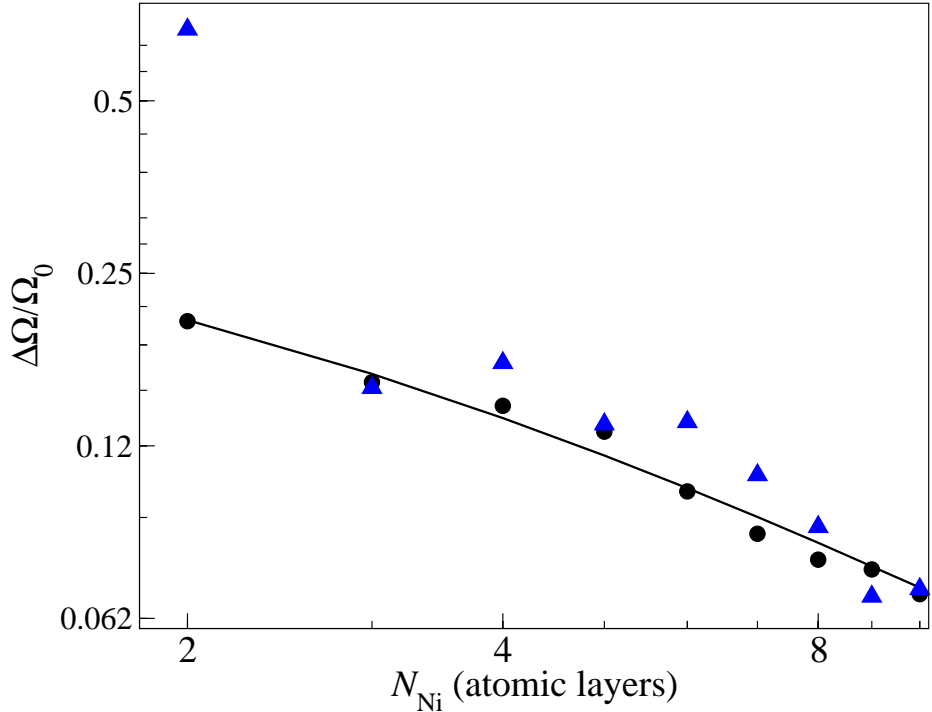}
\caption{Damping rate as a function of the Ni film thickness for Pd (solid circles) 
and non-enhanced Pd (solid triangles) substrates. The solid curve is a $1/(N+N_0)$ fitting of
the results for the enhanced substrate (solid circles).}
\label{FMR_NiPd}
\end{figure}

The effect of Stoner enhancement on spin pumping can be seen in Fig.~\ref{FMR_CoPd}. 
The solid circles are linewidth values extracted from FMR spectra calculated 
for $N$Co/Pd(001). $N$ has been varied from 1 to 10 and enhancement effects have been taken
into account within the first 10 layers of the Pd(001) substrate. 
We also show two fittings to the linewidth data for comparison:
the dashed black line is a $\Delta_0/N^\alpha$ fitting and the solid red line is a $\Delta_0/(N+N_0)$ fitting. 
$\Delta_0$, $\alpha$, and $N_0$ have been treated as adjustable parameters. Consider the $\Delta_0/N^\alpha$
fitting first: our results are best fitted to $\alpha\approx0.7$, which is smaller by 13\% than the
value for a non-enhanced Pd substrate. One way to interpret the effects of the enhancement is to
think about the polarisation of the substrate as an ``effective thickness'' added to the magnetic film.
In this case the linewidth \textit{versus} thickness curve should behave like 
$\Delta_0/(N+N_0)$, our second choice for a fitting. In this particular case we found the best fitting corresponds to $N_0\approx 1.4$. This suggests
that the Pd polarization acts as $\sim 1.4$ ``effective magnetic layers''. Going back to
the layer-by-layer distribution of magnetic moments presented in section~\ref{GSSection}, 
Fig.~\ref{magneticmomentsCoPd}, we notice that the integrated magnetic moment induced in Pd is $\sim 1.3$,
which is close to the average magnetic moment of the Co layers. Of course we did not expect $N_0$ to coincide
with the integrated magnetic moment induced in Pd, but the fact that both quantities are of the same order 
of magnitude is an indication that our interpretation makes sense. One further evidence in favour of
this interpretation is the fact that we find the effective thickness for Ni $N_0\approx 2.2$ (see Fig.~\ref{FMR_NiPd}). 
Since the magnetic moments of the Ni films are smaller than those of Co films, and the polarisation of the underlying
Pd is very similar in both cases, its role as an effective thickness is more pronounced in Ni. 

We show in the same figure the linewidths without Stoner enhancement in Pd, previously presented in 
Fig.~\ref{FMR_UPd_0}. Notice that there is an overall increase in linewidth when the Stoner enhancement
is turned on in the Pd substrate. This may be due to an increase in the interface transmissivity. 
Turning on Coulomb repulsion in the Pd substrate causes an exchange splitting between the majority 
and minority spin sub bands in Pd. That splitting decreases smoothly as one goes into the substrate, 
increasing the effective transmission coefficient of the interface. The effect is analogous to the
smoothing of a potential step. A large interface transmissivity means a larger spin current
pumped into the substrate and a larger damping rate to the FMR precession. It is also interesting to 
notice that the linewidth dependence on the Co thickness has a non-negligible
oscillatory component, which may be attributed to quantum interference effects.

The spin pumping rate for Ni films on Pd has a more complex behaviour than for Co
films, as can be seen in Fig.~\ref{FMR_NiPd}. With $U_\mathrm{Pd}=0$ it displays
strong oscillations as a function of $N$. While it is difficult to extract the rate of 
decay from a strongly oscillatory function with so few points, a $1/N$ curve fits
reasonably well our results. When Stoner enhancement is turned on the strong 
oscillations are partially suppressed and, as stated in the previous paragraph,
the results are well fitted to a $1/(N+N_0)$ curve (the solid curve in Fig.~\ref{FMR_NiPd}).

\subsection{Interlayer exchange coupling through Pd}  

There are many ways to estimate the magnitude of the exchange coupling between magnetic entities separated by
non magnetic metals.~\cite{Stiles1999322_IEC_Review} One widely used approach, that can be implemented in various forms, 
is to calculate energy differences between ferromagnetic and antiferromagnetic configurations. A more subtle approach 
is to calculate the total energy variation due to small deviations from either ferromagnetic or antiferromagnetic 
configurations.~\cite{Liechtenstein198765} This is what we refer to as the ``static'' exchange coupling. Both approaches 
rely on the assumption that the magnetic moments carried by the magnetic entities may be treated 
as  macrospins and their energy can be expressed as a function of their relative orientation. 



There is one way to estimate the exchange interaction energy between magnetic moments that does not depend on the 
kind of medium they are embedded into. This theoretical technique is identical in principle to the experimental
determination of IEC using ferromagnetic resonance experiments. Once the magnetic moments are perturbed from their 
equilibrium configuration their
precessional motion can be decomposed into normal modes, from which it is possible to extract the intensity of the 
interaction between them.~\cite{SpinPumpingTheory} It is simple to show, for example, that two magnetic moments
interacting via a Heisenberg-like term will have two normal modes, one in-phase (the so-called acoustic mode)
and one 180$^\circ$ out of phase (the ``optical'' mode). The difference between the energies of these two modes
is proportional to the strength of the coupling between the spins.

\begin{figure}
\includegraphics[width=0.8\textwidth]{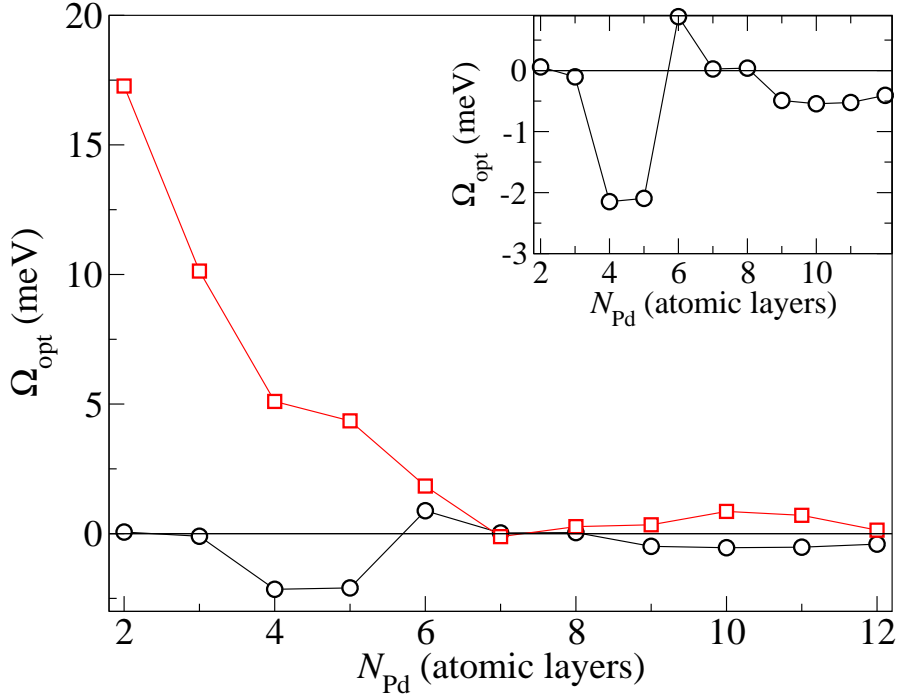}
\caption{Interlayer exchange coupling between two Co films (with two atomic layers each) separated
by $N_\mathrm{Pd}$ Pd atomic layers measured by the frequency of the optical 
FMR mode (squares). For comparison we show the same IEC calculated without exchange enhancement
within the Pd spacer (open circles). For clarity, due to the magnitude mismatch between the two results,
we repeat the IEC in the absence of Stoner enhancement
in the inset.}
\label{J_CoPd}
\end{figure}

\begin{figure}
\includegraphics[width=0.8\textwidth]{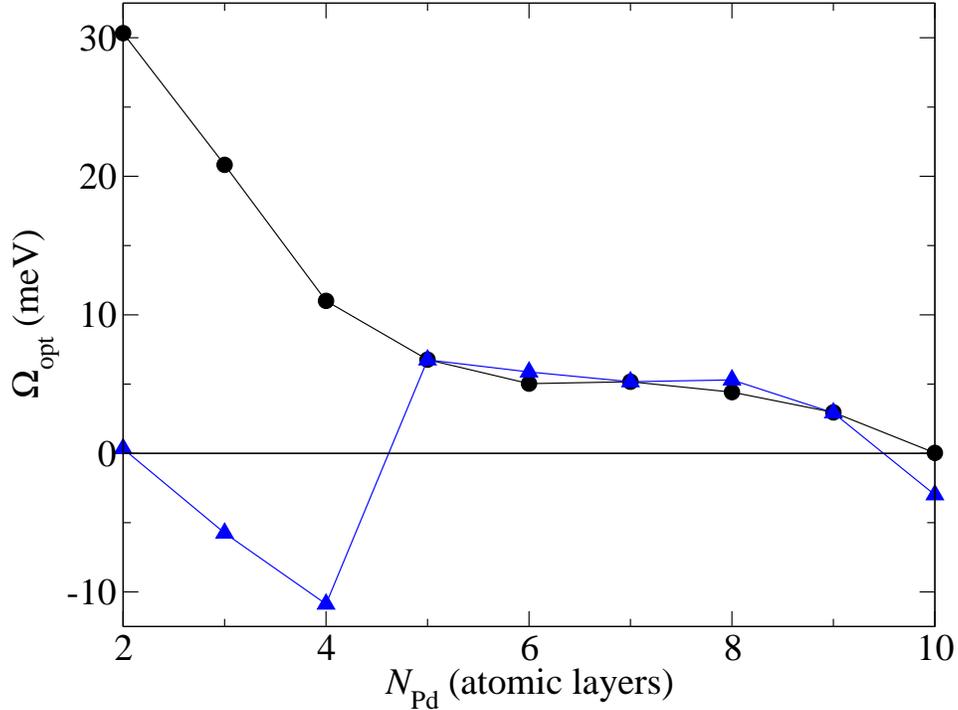}
\caption{Frequency of the optical (out-of-phase) precession mode of two Ni films (with two atomic layers each) separated
by $N_\mathrm{Pd}$ atomic layers.}
\label{J_NiPd}
\end{figure}

To investigate the effects of Coulomb repulsion on the IEC between magnetic ultrathin films
separated by an interacting (but nonmagnetic) medium we calculated the normal modes of trilayers formed by two 
ferromagnetic layers separated by a Pd slab. The results of our calculations are presented in Figs.~\ref{J_CoPd} and \ref{J_NiPd}. We only investigated small
Pd thicknesses because these calculations are extremely demanding computationally for thick spacers. In 
Fig.~\ref{J_CoPd} we present the frequency of the ``optical'' (out-of-phase) mode of two Co films separated by a $N$-atomic-layers thick Pd spacer taking Stoner enhancement into account. Even in this pre-asymptotic region of relatively small spacer thicknesses it is expected that the exchange coupling mediated by an independent electron system (for instance, typical non-magnetic metals) would display an oscillatory behavior. Our results indicate that the oscillations are suppressed by the very large Stoner enhancement in Pd. To confirm that this is not simply a special feature of the electronic structure of Pd we repeated the calculations with the same electronic structure but without Stoner enhancement. These results are shown in the inset of Fig.~\ref{J_CoPd}. Two differences are readily
noticed: in the absence of Stoner enhancement the exchange coupling is one order of magnitude smaller and there are clear oscillations.

Similar behaviour has been observed experimentally in Fe(001)/Pd(001)/Fe(001) multilayers.~\cite{FePdFe_experiment}
The suppression of the oscillatory behaviour is the most striking qualitative feature of the experimental results.

For Ni films there is a substantial change in the exchange coupling due to Stoner enhancement for small Pd 
thicknesses, as seen in Fig.~\ref{J_NiPd}. At large distances, however, the coupling seems to be little 
affected by the presence of Stoner enhancement within the Pd substrate. It is also interesting that the
strength of the exchange coupling is considerably larger between Ni films than between Co films
when Stoner enhancement is not active in the Pd spacer layer. This may be due to the large spin polarisability
of Pd even in the absence of Stoner enhancement. Since the magnetic moment of Ni is much smaller than that
of Co, even the small polarisation of non-enhanced Pd is enough to affect the exchange coupling at large
distances. Notice that at short distances the exchange coupling between Ni layers in the absence of Stoner 
enhancement would be antiferromagnetic. The Stoner enhancement in Pd changes the situation dramatically, since
Pd polarisation follows the neighbouring Ni magnetic moments and nearest-neighbor Pd-Pd exchange coupling is 
ferromagnetic. At large distances there is a relatively thick layer of Pd with very small magnetic moment in 
between the polarised Pd layers, which results in an effective system similar to the non-enhanced Pd spacer
sandwiched by Ni films.

\subsection{Dynamic exchange coupling through Pd}

A dynamic coupling between magnetic unities embedded into non-magnetic metals may be
active even when the IEC between them is negligible.~\cite{DynCouplGeneral,DynCouplCN,SpinPumpingTheory} 
This dynamic coupling is a consequence
of the spin current pumped by the magnetic unities when their magnetisations are perturbed dynamically
from their equilibrium configuration.\cite{Tserkovnyak_RMP2005} The perturbation may take the form of 
a magnetic field  pulse applied transverse to the equilibrium direction of the magnetisation, 
or a continuously applied transverse
field with harmonic time dependence, for example. One way to measure the intensity of such dynamic coupling
is through the non-local susceptibility $\chi_{AB}$, where $A$ and $B$ represent two distinct magnetic unities.
Schematically, if a transverse magnetic field $h^\perp_B$ is applied to unity $B$, $\chi_{AB}h^\perp_B$
is the transverse component acquired by the magnetisation of unit $A$. Since the number of spacer
layers may affect also the local response $\chi_{AA}$ we chose the ratio 
$\mathrm{Im}\chi_{AB}/\mathrm{Im}\chi_{AA}$ as a measure of the dynamic coupling between films $A$ and $B$.

\begin{figure}
\includegraphics[width=0.8\textwidth]{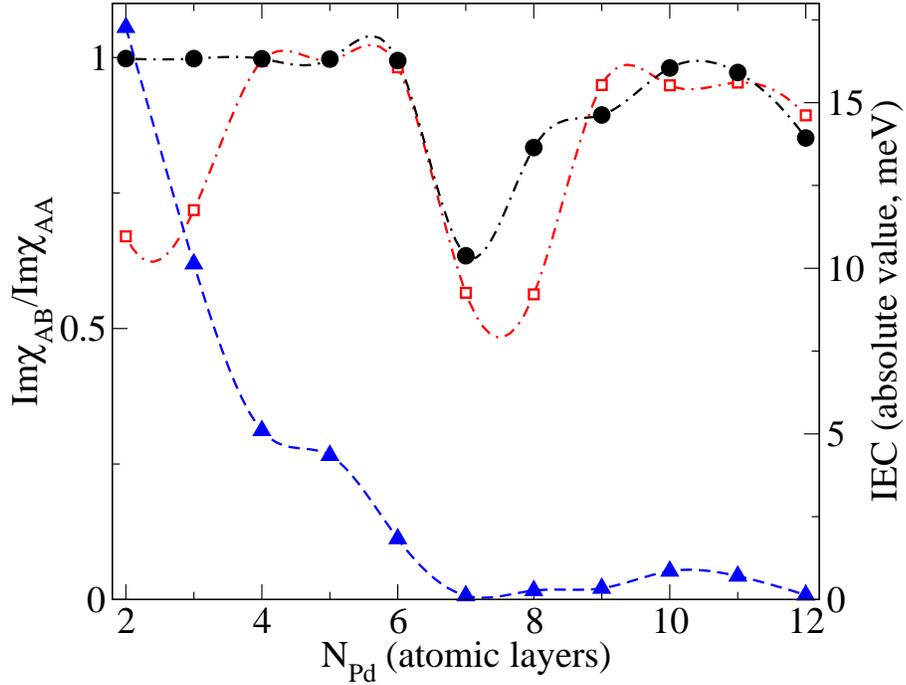}
\caption{Dynamic coupling between two Co films (with two atomic layers each) separated
by $N_\mathrm{Pd}$ atomic layers. The results represented by solid circles where obtained
with Stoner enhancement turned off within the Pd spacer. The dashed and dot-dashed lines are 
meant as guides to the eye only. The absolute value of the static IEC (solid triangles) 
is also shown for comparison (scale on the right).}
\label{DynCoupl_CoPd}
\end{figure}

We calculated the dynamic coupling between two Co films separated by Pd layers with varying
thickness $N_\mathrm{Pd}$ via the non-local transverse susceptibility defined in Eq.~\ref{qparchipm}.
We set $\vec{Q}_\parallel=0$ and $\Omega$ was chosen as the resonance frequency for the ``acoustic'' 
(in-phase) mode. The results are shown in Fig.~\ref{DynCoupl_CoPd}. To assess the effect of Stoner
enhancement we performed calculations in which the effective Coulomb interaction was set to zero
within the Pd spacer layer (open squares in Fig.~\ref{DynCoupl_CoPd}). The dynamic coupling in the absence
of Stoner enhancement oscillates as a function of $N_\mathrm{Pd}$ with a constant amplitude.
When Stoner enhancement is turned on the oscillations are strongly suppressed for $N_\mathrm{Pd}\le 6$.
For such thicknessess the whole Pd spacer is strongly spin-polarized. For $N_\mathrm{Pd}> 6$
there are atomic planes of Pd with negligible polarization close to the center of the spacer layer,
which restore the oscillatory behavior of the dynamic coupling. 

It is also interesting to remind ourselves about the behavior of the static IEC we calculated previously.
We plot the static IEC as solid triangles in Fig.~\ref{DynCoupl_CoPd}. Notice that it decays relatively fast
with $N_\mathrm{Pd}$, being very close to zero for $N_\mathrm{Pd}\ge 7$, whereas the dynamic coupling 
is clearly finite for the entire range of $N_\mathrm{Pd}$ considered. In fact, the range of the dynamic 
coupling is infinite within our model, since we do not consider any mechanism that could dissipate
the spin current emitted by the pumped Co film. This is equivalent to saying that the range of the dynamic 
coupling should only be limited by the spin diffusion length.

\begin{figure}
\includegraphics[width=0.8\textwidth]{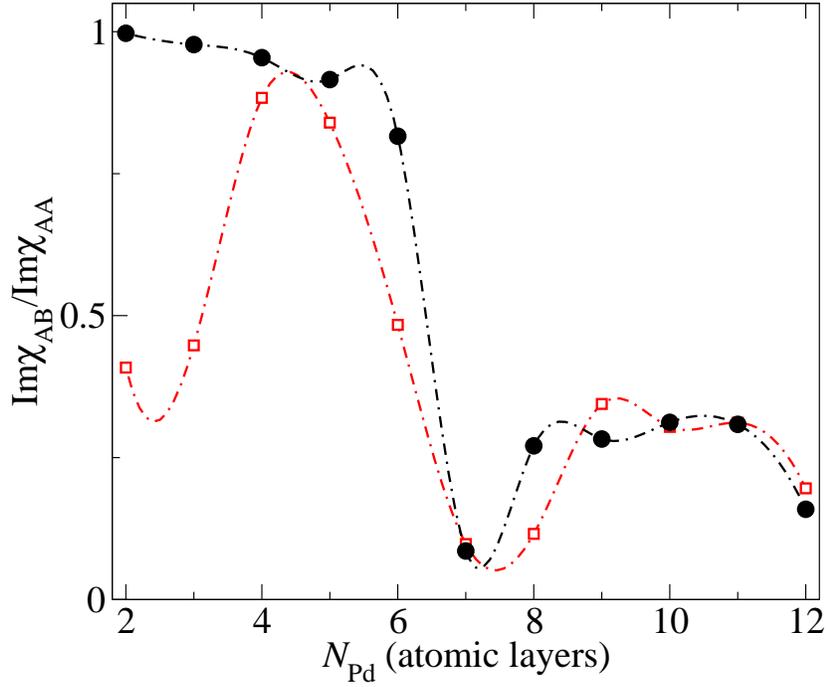}
\caption{Dynamic coupling between two Co films (with two atomic layers each) separated
by $N_\mathrm{Pd}$ atomic layers for a finite wave vector of the pumping field, 
$\vec{Q}=\frac{0.44}{a_0}\hat{i}$. The results represented by solid circles where obtained
with Stoner enhancement turned off within the Pd spacer. The dashed and dot-dashed lines are 
meant as guides to the eye only.}
\label{DynCoupl_CoPd_Q0_07}
\end{figure}

The above discussion about the infinite range of dynamic coupling applies to cases where the pumping
field is uniform in the direction parallel to the ferromagnetic layers. In that case the precession
mode being excited is the uniform ($\vec{Q}=0$)  mode and rotation symmetry in spin space prohibits 
decaying of this mode into uncorrelated electron-hole pairs (the so-called Stoner excitations). 
If, however, the pumping field has a finite wave vector, decaying of the spin wave into Stoner excitations 
is allowed and the dynamic coupling should acquire a finite range as a function of $N_\mathrm{Pd}$.
This decaying is shown clearly in Fig.~\ref{DynCoupl_CoPd_Q0_07}, which depicts results for 
the dynamic coupling $\mathrm{Im}\chi_{AB}/\mathrm{Im}\chi_{AA}$
calculated for a finite wave vector $\vec{Q}=\frac{0.44}{a_0}\hat{\imath}$. 
As above, we present results in the absence of Stoner enhancement (open circles) for comparison.
The decaying is relatively slow for small Pd thicknesses, indicating that the strong polarization
of the substrate suppresses partially the decaying into Stoner modes. This is expected since spin 
polarization reduces the density of Stoner modes. 

\section{Concluding Remarks}
\label{concludingsection}

We have investigated the influence of Stoner enhancement on spin pumping and exchange coupling,
dynamic and static, through Pd substrates. We used realistic models to describe the electronic
structures of both magnetic films and substrate. Our formalism to study spin pumping and
interlayer exchange coupling, static and dynamic, is based on the calculation of the dynamic
response of the system to time dependent magnetic fields transverse to the equilibrium direction of
the system's magnetization. We
found that Stoner enhancement has a small but non-negligible effect on the spin pumping rate
at zero transverse wave vector and without spin-orbit coupling. However, Stoner enhancement
within Pd can strongly affect the behaviour of IEC between Co and Ni films: it suppresses the
oscillations of the IEC as a function of spacer thickness, and increases substantially the 
strength of the IEC. Stoner enhancement also suppresses the oscillations of the dynamic 
interlayer coupling both at zero and finite wave vector. Investigations of the combined effect of
Stoner enhancement and spin-orbit coupling are underway.

\begin{acknowledgments}
The authors gratefully acknowledge partial financial support from the Brazilian agencies 
CNPq, CAPES and FAPERJ.
\end{acknowledgments}

\bibliography{referencias}

\begin{thebibliography}{29}
\expandafter\ifx\csname natexlab\endcsname\relax\def\natexlab#1{#1}\fi
\expandafter\ifx\csname bibnamefont\endcsname\relax
  \def\bibnamefont#1{#1}\fi
\expandafter\ifx\csname bibfnamefont\endcsname\relax
  \def\bibfnamefont#1{#1}\fi
\expandafter\ifx\csname citenamefont\endcsname\relax
  \def\citenamefont#1{#1}\fi
\expandafter\ifx\csname url\endcsname\relax
  \def\url#1{\texttt{#1}}\fi
\expandafter\ifx\csname urlprefix\endcsname\relax\def\urlprefix{URL }\fi
\providecommand{\bibinfo}[2]{#2}
\providecommand{\eprint}[2][]{\url{#2}}

\bibitem[{\citenamefont{Lederer and Mills}(1968)}]{LedererMills_NiPd}
\bibinfo{author}{\bibfnamefont{P.}~\bibnamefont{Lederer}} \bibnamefont{and}
  \bibinfo{author}{\bibfnamefont{D.~L.} \bibnamefont{Mills}},
  \bibinfo{journal}{Phys. Rev.} \textbf{\bibinfo{volume}{165}},
  \bibinfo{pages}{837} (\bibinfo{year}{1968}),
  \urlprefix\url{http://link.aps.org/doi/10.1103/PhysRev.165.837}.

\bibitem[{\citenamefont{Doubble et~al.}(2010)\citenamefont{Doubble, Hayden,
  Dai, Mook, Thompson, and Frost}}]{paramagnon_observation}
\bibinfo{author}{\bibfnamefont{R.}~\bibnamefont{Doubble}},
  \bibinfo{author}{\bibfnamefont{S.~M.} \bibnamefont{Hayden}},
  \bibinfo{author}{\bibfnamefont{P.}~\bibnamefont{Dai}},
  \bibinfo{author}{\bibfnamefont{H.~A.} \bibnamefont{Mook}},
  \bibinfo{author}{\bibfnamefont{J.~R.} \bibnamefont{Thompson}},
  \bibnamefont{and} \bibinfo{author}{\bibfnamefont{C.~D.} \bibnamefont{Frost}},
  \bibinfo{journal}{Phys. Rev. Lett.} \textbf{\bibinfo{volume}{105}},
  \bibinfo{pages}{027207} (\bibinfo{year}{2010}),
  \urlprefix\url{http://link.aps.org/doi/10.1103/PhysRevLett.105.027207}.

\bibitem[{\citenamefont{Alexandre et~al.}(2006)\citenamefont{Alexandre,
  Mattesini, Soler, and Yndurain}}]{PhysRevLett.96.079701}
\bibinfo{author}{\bibfnamefont{S.~S.} \bibnamefont{Alexandre}},
  \bibinfo{author}{\bibfnamefont{M.}~\bibnamefont{Mattesini}},
  \bibinfo{author}{\bibfnamefont{J.~M.} \bibnamefont{Soler}}, \bibnamefont{and}
  \bibinfo{author}{\bibfnamefont{F.}~\bibnamefont{Yndurain}},
  \bibinfo{journal}{Phys. Rev. Lett.} \textbf{\bibinfo{volume}{96}},
  \bibinfo{pages}{079701} (\bibinfo{year}{2006}),
  \urlprefix\url{http://link.aps.org/doi/10.1103/PhysRevLett.96.079701}.

\bibitem[{\citenamefont{Costa et~al.}(2010)\citenamefont{Costa, Muniz, Lounis,
  Klautau, and Mills}}]{FeW110_SOC_Theory}
\bibinfo{author}{\bibfnamefont{A.~T.} \bibnamefont{Costa}},
  \bibinfo{author}{\bibfnamefont{R.~B.} \bibnamefont{Muniz}},
  \bibinfo{author}{\bibfnamefont{S.}~\bibnamefont{Lounis}},
  \bibinfo{author}{\bibfnamefont{A.~B.} \bibnamefont{Klautau}},
  \bibnamefont{and} \bibinfo{author}{\bibfnamefont{D.~L.} \bibnamefont{Mills}},
  \bibinfo{journal}{Phys. Rev. B} \textbf{\bibinfo{volume}{82}},
  \bibinfo{pages}{014428} (\bibinfo{year}{2010}),
  \urlprefix\url{http://link.aps.org/doi/10.1103/PhysRevB.82.014428}.

\bibitem[{\citenamefont{Zakeri et~al.}(2010)\citenamefont{Zakeri, Zhang,
  Prokop, Chuang, Sakr, Tang, and Kirschner}}]{FeW110_SOC_Experiment_1}
\bibinfo{author}{\bibfnamefont{K.}~\bibnamefont{Zakeri}},
  \bibinfo{author}{\bibfnamefont{Y.}~\bibnamefont{Zhang}},
  \bibinfo{author}{\bibfnamefont{J.}~\bibnamefont{Prokop}},
  \bibinfo{author}{\bibfnamefont{T.-H.} \bibnamefont{Chuang}},
  \bibinfo{author}{\bibfnamefont{N.}~\bibnamefont{Sakr}},
  \bibinfo{author}{\bibfnamefont{W.~X.} \bibnamefont{Tang}}, \bibnamefont{and}
  \bibinfo{author}{\bibfnamefont{J.}~\bibnamefont{Kirschner}},
  \bibinfo{journal}{Phys. Rev. Lett.} \textbf{\bibinfo{volume}{104}},
  \bibinfo{pages}{137203} (\bibinfo{year}{2010}),
  \urlprefix\url{http://link.aps.org/doi/10.1103/PhysRevLett.104.137203}.

\bibitem[{\citenamefont{Khajetoorians et~al.}(2011)\citenamefont{Khajetoorians,
  Lounis, Chilian, Costa, Zhou, Mills, Wiebe, and Wiesendanger}}]{FeCu111_STS}
\bibinfo{author}{\bibfnamefont{A.~A.} \bibnamefont{Khajetoorians}},
  \bibinfo{author}{\bibfnamefont{S.}~\bibnamefont{Lounis}},
  \bibinfo{author}{\bibfnamefont{B.}~\bibnamefont{Chilian}},
  \bibinfo{author}{\bibfnamefont{A.~T.} \bibnamefont{Costa}},
  \bibinfo{author}{\bibfnamefont{L.}~\bibnamefont{Zhou}},
  \bibinfo{author}{\bibfnamefont{D.~L.} \bibnamefont{Mills}},
  \bibinfo{author}{\bibfnamefont{J.}~\bibnamefont{Wiebe}}, \bibnamefont{and}
  \bibinfo{author}{\bibfnamefont{R.}~\bibnamefont{Wiesendanger}},
  \bibinfo{journal}{Phys. Rev. Lett.} \textbf{\bibinfo{volume}{106}},
  \bibinfo{pages}{037205} (\bibinfo{year}{2011}),
  \urlprefix\url{http://link.aps.org/doi/10.1103/PhysRevLett.106.037205}.

\bibitem[{\citenamefont{Chilian et~al.}(2011)\citenamefont{Chilian,
  Khajetoorians, Lounis, Costa, Mills, Wiebe, and Wiesendanger}}]{FeAg111_STS}
\bibinfo{author}{\bibfnamefont{B.}~\bibnamefont{Chilian}},
  \bibinfo{author}{\bibfnamefont{A.~A.} \bibnamefont{Khajetoorians}},
  \bibinfo{author}{\bibfnamefont{S.}~\bibnamefont{Lounis}},
  \bibinfo{author}{\bibfnamefont{A.~T.} \bibnamefont{Costa}},
  \bibinfo{author}{\bibfnamefont{D.~L.} \bibnamefont{Mills}},
  \bibinfo{author}{\bibfnamefont{J.}~\bibnamefont{Wiebe}}, \bibnamefont{and}
  \bibinfo{author}{\bibfnamefont{R.}~\bibnamefont{Wiesendanger}},
  \bibinfo{journal}{Phys. Rev. B} \textbf{\bibinfo{volume}{84}},
  \bibinfo{pages}{212401} (\bibinfo{year}{2011}),
  \urlprefix\url{http://link.aps.org/doi/10.1103/PhysRevB.84.212401}.

\bibitem[{\citenamefont{Foros et~al.}(2005)\citenamefont{Foros, Woltersdorf,
  Heinrich, and Brataas}}]{heinrich_Pd}
\bibinfo{author}{\bibfnamefont{J.}~\bibnamefont{Foros}},
  \bibinfo{author}{\bibfnamefont{G.}~\bibnamefont{Woltersdorf}},
  \bibinfo{author}{\bibfnamefont{B.}~\bibnamefont{Heinrich}}, \bibnamefont{and}
  \bibinfo{author}{\bibfnamefont{A.}~\bibnamefont{Brataas}},
  \bibinfo{journal}{Journal of Applied Physics} \textbf{\bibinfo{volume}{97}},
  \bibinfo{eid}{10A714} (pages~\bibinfo{numpages}{3}) (\bibinfo{year}{2005}),
  \urlprefix\url{http://link.aip.org/link/?JAP/97/10A714/1}.

\bibitem[{\citenamefont{Takahashi}(1997)}]{takahashi_FePdFe}
\bibinfo{author}{\bibfnamefont{Y.}~\bibnamefont{Takahashi}},
  \bibinfo{journal}{Phys. Rev. B} \textbf{\bibinfo{volume}{56}},
  \bibinfo{pages}{8175} (\bibinfo{year}{1997}),
  \urlprefix\url{http://link.aps.org/doi/10.1103/PhysRevB.56.8175}.

\bibitem[{\citenamefont{Slonczewski}(1999)}]{Slonczewski1999L261}
\bibinfo{author}{\bibfnamefont{J.}~\bibnamefont{Slonczewski}},
  \bibinfo{journal}{Journal of Magnetism and Magnetic Materials}
  \textbf{\bibinfo{volume}{195}}, \bibinfo{pages}{L261 }
  (\bibinfo{year}{1999}), ISSN \bibinfo{issn}{0304-8853},
  \urlprefix\url{http://www.sciencedirect.com/science/article/pii/S0304885399000438}.

\bibitem[{\citenamefont{Berger}(1996)}]{berger_PhysRevB.54.9353}
\bibinfo{author}{\bibfnamefont{L.}~\bibnamefont{Berger}},
  \bibinfo{journal}{Phys. Rev. B} \textbf{\bibinfo{volume}{54}},
  \bibinfo{pages}{9353} (\bibinfo{year}{1996}),
  \urlprefix\url{http://link.aps.org/doi/10.1103/PhysRevB.54.9353}.

\bibitem[{\citenamefont{Tserkovnyak et~al.}(2005)\citenamefont{Tserkovnyak,
  Brataas, Bauer, and Halperin}}]{Tserkovnyak_RMP2005}
\bibinfo{author}{\bibfnamefont{Y.}~\bibnamefont{Tserkovnyak}},
  \bibinfo{author}{\bibfnamefont{A.}~\bibnamefont{Brataas}},
  \bibinfo{author}{\bibfnamefont{G.~E.~W.} \bibnamefont{Bauer}},
  \bibnamefont{and} \bibinfo{author}{\bibfnamefont{B.~I.}
  \bibnamefont{Halperin}}, \bibinfo{journal}{Rev. Mod. Phys.}
  \textbf{\bibinfo{volume}{77}}, \bibinfo{pages}{1375} (\bibinfo{year}{2005}),
  \urlprefix\url{http://link.aps.org/doi/10.1103/RevModPhys.77.1375}.

\bibitem[{\citenamefont{Barbosa et~al.}(2001)\citenamefont{Barbosa, Muniz,
  Costa, and Mathon}}]{SW_1band}
\bibinfo{author}{\bibfnamefont{L.~H.~M.} \bibnamefont{Barbosa}},
  \bibinfo{author}{\bibfnamefont{R.~B.} \bibnamefont{Muniz}},
  \bibinfo{author}{\bibfnamefont{A.~T.} \bibnamefont{Costa}}, \bibnamefont{and}
  \bibinfo{author}{\bibfnamefont{J.}~\bibnamefont{Mathon}},
  \bibinfo{journal}{Phys. Rev. B} \textbf{\bibinfo{volume}{63}},
  \bibinfo{pages}{174401} (\bibinfo{year}{2001}),
  \urlprefix\url{http://link.aps.org/doi/10.1103/PhysRevB.63.174401}.

\bibitem[{\citenamefont{Costa et~al.}(2003)\citenamefont{Costa, Muniz, and
  Mills}}]{SW_Fe110_old1}
\bibinfo{author}{\bibfnamefont{A.~T.} \bibnamefont{Costa}},
  \bibinfo{author}{\bibfnamefont{R.~B.} \bibnamefont{Muniz}}, \bibnamefont{and}
  \bibinfo{author}{\bibfnamefont{D.~L.} \bibnamefont{Mills}},
  \bibinfo{journal}{Phys. Rev. B} \textbf{\bibinfo{volume}{68}},
  \bibinfo{pages}{224435} (\bibinfo{year}{2003}),
  \urlprefix\url{http://link.aps.org/doi/10.1103/PhysRevB.68.224435}.

\bibitem[{\citenamefont{Muniz and Mills}(2002)}]{SW_Fe110_old0}
\bibinfo{author}{\bibfnamefont{R.~B.} \bibnamefont{Muniz}} \bibnamefont{and}
  \bibinfo{author}{\bibfnamefont{D.~L.} \bibnamefont{Mills}},
  \bibinfo{journal}{Phys. Rev. B} \textbf{\bibinfo{volume}{66}},
  \bibinfo{pages}{174417} (\bibinfo{year}{2002}),
  \urlprefix\url{http://link.aps.org/doi/10.1103/PhysRevB.66.174417}.

\bibitem[{\citenamefont{Costa et~al.}(2006)\citenamefont{Costa, Bechara~Muniz,
  and Mills}}]{SpinPumpingTheory}
\bibinfo{author}{\bibfnamefont{A.~T.} \bibnamefont{Costa}},
  \bibinfo{author}{\bibfnamefont{R.}~\bibnamefont{Bechara~Muniz}},
  \bibnamefont{and} \bibinfo{author}{\bibfnamefont{D.~L.} \bibnamefont{Mills}},
  \bibinfo{journal}{Phys. Rev. B} \textbf{\bibinfo{volume}{73}},
  \bibinfo{pages}{054426} (\bibinfo{year}{2006}),
  \urlprefix\url{http://link.aps.org/doi/10.1103/PhysRevB.73.054426}.

\bibitem[{\citenamefont{Costa et~al.}(2008{\natexlab{a}})\citenamefont{Costa,
  Muniz, Ferreira, and Mills}}]{DynCouplGeneral}
\bibinfo{author}{\bibfnamefont{A.~T.} \bibnamefont{Costa}},
  \bibinfo{author}{\bibfnamefont{R.~B.} \bibnamefont{Muniz}},
  \bibinfo{author}{\bibfnamefont{M.~S.} \bibnamefont{Ferreira}},
  \bibnamefont{and} \bibinfo{author}{\bibfnamefont{D.~L.} \bibnamefont{Mills}},
  \bibinfo{journal}{Phys. Rev. B} \textbf{\bibinfo{volume}{78}},
  \bibinfo{pages}{214403} (\bibinfo{year}{2008}{\natexlab{a}}),
  \urlprefix\url{http://link.aps.org/doi/10.1103/PhysRevB.78.214403}.

\bibitem[{\citenamefont{Costa et~al.}(2008{\natexlab{b}})\citenamefont{Costa,
  Muniz, and Ferreira}}]{DynCouplCN}
\bibinfo{author}{\bibfnamefont{A.~T.} \bibnamefont{Costa}},
  \bibinfo{author}{\bibfnamefont{R.~B.} \bibnamefont{Muniz}}, \bibnamefont{and}
  \bibinfo{author}{\bibfnamefont{M.~S.} \bibnamefont{Ferreira}},
  \bibinfo{journal}{New Journal of Physics} \textbf{\bibinfo{volume}{10}},
  \bibinfo{pages}{063008} (\bibinfo{year}{2008}{\natexlab{b}}),
  \urlprefix\url{http://stacks.iop.org/1367-2630/10/i=6/a=063008}.

\bibitem[{\citenamefont{Giovannini et~al.}(1964)\citenamefont{Giovannini,
  Peter, and Schrieffer}}]{schrieffer1964}
\bibinfo{author}{\bibfnamefont{B.}~\bibnamefont{Giovannini}},
  \bibinfo{author}{\bibfnamefont{M.}~\bibnamefont{Peter}}, \bibnamefont{and}
  \bibinfo{author}{\bibfnamefont{J.~R.} \bibnamefont{Schrieffer}},
  \bibinfo{journal}{Phys. Rev. Lett.} \textbf{\bibinfo{volume}{12}},
  \bibinfo{pages}{736} (\bibinfo{year}{1964}),
  \urlprefix\url{http://link.aps.org/doi/10.1103/PhysRevLett.12.736}.

\bibitem[{\citenamefont{Doniach and Engelsberg}(1966)}]{DoniachEngelsberg1966}
\bibinfo{author}{\bibfnamefont{S.}~\bibnamefont{Doniach}} \bibnamefont{and}
  \bibinfo{author}{\bibfnamefont{S.}~\bibnamefont{Engelsberg}},
  \bibinfo{journal}{Phys. Rev. Lett.} \textbf{\bibinfo{volume}{17}},
  \bibinfo{pages}{750} (\bibinfo{year}{1966}),
  \urlprefix\url{http://link.aps.org/doi/10.1103/PhysRevLett.17.750}.

\bibitem[{\citenamefont{Low and Holden}(1966)}]{GiantMM_Pd}
\bibinfo{author}{\bibfnamefont{G.~G.} \bibnamefont{Low}} \bibnamefont{and}
  \bibinfo{author}{\bibfnamefont{T.~M.} \bibnamefont{Holden}},
  \bibinfo{journal}{Proceedings of the Physical Society}
  \textbf{\bibinfo{volume}{89}}, \bibinfo{pages}{119} (\bibinfo{year}{1966}),
  \urlprefix\url{http://stacks.iop.org/0370-1328/89/i=1/a=318}.

\bibitem[{\citenamefont{Kresse and Hafner}(1994)}]{vasp1}
\bibinfo{author}{\bibfnamefont{G.}~\bibnamefont{Kresse}} \bibnamefont{and}
  \bibinfo{author}{\bibfnamefont{J.}~\bibnamefont{Hafner}},
  \bibinfo{journal}{\prb} \textbf{\bibinfo{volume}{49}}, \bibinfo{pages}{14251}
  (\bibinfo{year}{1994}).

\bibitem[{\citenamefont{Kresse and Furthmüller}(1996)}]{vasp2}
\bibinfo{author}{\bibfnamefont{G.}~\bibnamefont{Kresse}} \bibnamefont{and}
  \bibinfo{author}{\bibfnamefont{J.}~\bibnamefont{Furthmüller}},
  \bibinfo{journal}{Comput. Mat. Sci.} \textbf{\bibinfo{volume}{6}},
  \bibinfo{pages}{15} (\bibinfo{year}{1996}).

\bibitem[{\citenamefont{Perdew and Zunger}(1981)}]{perdewzunger}
\bibinfo{author}{\bibfnamefont{J.~P.} \bibnamefont{Perdew}} \bibnamefont{and}
  \bibinfo{author}{\bibfnamefont{A.}~\bibnamefont{Zunger}},
  \bibinfo{journal}{\prb} \textbf{\bibinfo{volume}{23}}, \bibinfo{pages}{5048}
  (\bibinfo{year}{1981}).

\bibitem[{\citenamefont{Kresse and Joubert}(1999)}]{paw_method}
\bibinfo{author}{\bibfnamefont{G.}~\bibnamefont{Kresse}} \bibnamefont{and}
  \bibinfo{author}{\bibfnamefont{D.}~\bibnamefont{Joubert}},
  \bibinfo{journal}{Phys. Rev. B} \textbf{\bibinfo{volume}{59}},
  \bibinfo{pages}{1758} (\bibinfo{year}{1999}),
  \urlprefix\url{http://link.aps.org/doi/10.1103/PhysRevB.59.1758}.

\bibitem[{\citenamefont{Monkhorst and Pack}(1976)}]{MonkhorstPack}
\bibinfo{author}{\bibfnamefont{H.~J.} \bibnamefont{Monkhorst}}
  \bibnamefont{and} \bibinfo{author}{\bibfnamefont{J.~D.} \bibnamefont{Pack}},
  \bibinfo{journal}{Phys. Rev. B} \textbf{\bibinfo{volume}{13}},
  \bibinfo{pages}{5188} (\bibinfo{year}{1976}),
  \urlprefix\url{http://link.aps.org/doi/10.1103/PhysRevB.13.5188}.

\bibitem[{\citenamefont{Stiles}(1999)}]{Stiles1999322_IEC_Review}
\bibinfo{author}{\bibfnamefont{M.}~\bibnamefont{Stiles}},
  \bibinfo{journal}{Journal of Magnetism and Magnetic Materials}
  \textbf{\bibinfo{volume}{200}}, \bibinfo{pages}{322 } (\bibinfo{year}{1999}),
  ISSN \bibinfo{issn}{0304-8853},
  \urlprefix\url{http://www.sciencedirect.com/science/article/pii/S0304885399003340}.

\bibitem[{\citenamefont{Liechtenstein et~al.}(1987)\citenamefont{Liechtenstein,
  Katsnelson, Antropov, and Gubanov}}]{Liechtenstein198765}
\bibinfo{author}{\bibfnamefont{A.}~\bibnamefont{Liechtenstein}},
  \bibinfo{author}{\bibfnamefont{M.}~\bibnamefont{Katsnelson}},
  \bibinfo{author}{\bibfnamefont{V.}~\bibnamefont{Antropov}}, \bibnamefont{and}
  \bibinfo{author}{\bibfnamefont{V.}~\bibnamefont{Gubanov}},
  \bibinfo{journal}{Journal of Magnetism and Magnetic Materials}
  \textbf{\bibinfo{volume}{67}}, \bibinfo{pages}{65 } (\bibinfo{year}{1987}),
  ISSN \bibinfo{issn}{0304-8853},
  \urlprefix\url{http://www.sciencedirect.com/science/article/pii/0304885387907219}.

\bibitem[{\citenamefont{Celinski et~al.}(1990)\citenamefont{Celinski, Heinrich,
  Cochran, Muir, Arrott, and Kirschner}}]{FePdFe_experiment}
\bibinfo{author}{\bibfnamefont{Z.}~\bibnamefont{Celinski}},
  \bibinfo{author}{\bibfnamefont{B.}~\bibnamefont{Heinrich}},
  \bibinfo{author}{\bibfnamefont{J.~F.} \bibnamefont{Cochran}},
  \bibinfo{author}{\bibfnamefont{W.~B.} \bibnamefont{Muir}},
  \bibinfo{author}{\bibfnamefont{A.~S.} \bibnamefont{Arrott}},
  \bibnamefont{and}
  \bibinfo{author}{\bibfnamefont{J.}~\bibnamefont{Kirschner}},
  \bibinfo{journal}{Phys. Rev. Lett.} \textbf{\bibinfo{volume}{65}},
  \bibinfo{pages}{1156} (\bibinfo{year}{1990}),
  \urlprefix\url{http://link.aps.org/doi/10.1103/PhysRevLett.65.1156}.

\end{thebibliography}

\end{document}